\documentclass[rnote]{aa}
\usepackage{graphicx}
\usepackage[sectionbib,authoryear]{natbib}
\usepackage{times}
\usepackage{graphicx} 

\def\simge{\mathrel{    \rlap{\raise 0.511ex \hbox{$>$}}{\lower
0.511ex \hbox{$\sim$}}}}
\def\simle{\mathrel{    \rlap{\raise 0.511ex \hbox{$<$}}{\lower
0.511ex \hbox{$\sim$}}}}

\newcommand{\gr}{$\gamma$-ray \,}

\newcommand{\hess}{H.E.S.S.}
\newcommand{\grs}{$\gamma$-rays \,}
\usepackage[dvips]{color}
\usepackage[normalem]{ulem}
\usepackage{fancybox}

\begin{document}

\title{Inverse Compton gamma-ray models for remnants of Galactic type Ia
supernovae ?}


\authorrunning{V\"olk et al.}  
\titlerunning{IC \gr\ models for type Ia SNRs} 

\author{ H.J. V\"olk \inst{1}
         \and 
          L.T. Ksenofontov 
         \inst{2}
          \and 
          E.G. Berezhko 
          \inst{2}
}

\institute{Max Planck Institut f\"ur Kernphysik,
                Postfach 103980, D-69029 Heidelberg, Germany\\
              \email{Heinrich.Voelk@mpi-hd.mpg.de}
           \and
            Yu.G. Shafer Institute of Cosmophysical Research and Aeronomy,
                     31 Lenin Ave., 677980 Yakutsk, Russia\\
               \email{ksenofon@ikfia.ysn.ru}
                \email{berezhko@ikfia.ysn.ru}         
}
	     
\offprints{H.J. V\"olk}

\date{Received month day, year; accepted month day, year}

\abstract {} {We theoretically and phenomenologically investigate the
question whether the \gr emission from the remnants of the type Ia
supernovae SN 1006, Tycho's SN and Kepler's SN can be the result of
electron acceleration alone.}  
{The observed synchrotron spectra of the three remnants are used to
determine the average momentum distribution of nonthermal electrons as
a function of the assumed magnetic field strength. Then the inverse
Compton emission spectrum in the Cosmic Microwave Background photon
field is calculated and compared with the existing upper limits for
the very high energy \gr flux from these sources.}
{It is shown that the expected interstellar magnetic fields
substantially overpredict even these \gr upper limits. Only rather
strongly amplified magnetic fields could be compatible with such low
\gr fluxes. However this would require a strong component of
accelerated nuclear particles whose energy density substantially
exceeds that of the synchrotron electrons, compatible with existing
theoretical acceleration models for nuclear particles and electrons.}
{Even though the quantitative arguments are simplistic, they appear to
eliminate simplistic phenomenological claims in favor of a
inverse Compton \gr scenario for these sources.}

\keywords{(ISM:)cosmic rays -- acceleration of particles -- shock
waves -- supernovae individual (SN 1006, Tycho's SN, Kepler's SN) --
radiation mechanisms:non-thermal -- gamma-rays:theory}

\maketitle

%

\section{Introduction}
The question, whether the very high energy (VHE) ($E_\gamma>100$~GeV)
\gr emission of the Galactic supernova remnants (SNRs) implies a
sufficiently large production of nuclear
cosmic rays (CRs) -- of the same order as that required to replenish
the Galactic CRs -- is one of the key problems addressed by \gr
astronomy. There are two ways to deal with this question in the
investigation of an individual SNR. 

The first approach is a theoretical one. It uses a nonlinear
combination of gas dynamics (or eventually magnetohydrodynamics) for
the thermal gas/plasma and kinetic transport theory for the
collisionless, nonthermal relativistic particle component that is
coupled with the plasma physics of the electromagnetic field
fluctuations which scatter these particles. In the environment of the
collisionless shock wave of a supernova explosion this allows the
description of diffusive shock acceleration of the energetic particles
which are originally extracted from the thermal gas and thus {\it
injected} into the acceleration process. Since the field fluctuations
are excited by the accelerating particles themselves and since the
pressure of these particles (typically comparable with the thermal
pressure) is reacting back on the thermal plasma, this strongly
coupled system becomes a complex problem of
nonlinear dynamics, not only for the charged particle components but
also for the electromagnetic field and its fluctuations.
Models suggest that a sizeable fraction of the entire
hydrodynamic explosion energy will be transformed into energetic
particle energy. This
suggests that the SNRs are indeed the sources of the Galactic CRs.

Both nuclear charged particles and electrons can be accelerated to
achieve nonthermal momentum distributions. The energetic electrons
show their presence through synchrotron emission from radio
frequencies to hard X-ray energies. They may also interact with
diffuse interstellar radiation field photons, like the
Cosmic Microwave Background, to produce high energy \grs
in inverse Compton (IC) collisions. The injection of electrons into
the acceleration process is however poorly understood
quantitatively. Even assuming that the electron momentum distribution
at high particle energies is only proportional to the total number of
electrons injected per unit time at the shock, the amplitude factor of
the electron momentum distribution is therefore not known from
theory. It is typically inferred from the measured synchrotron
spectrum produced by the accelerated electrons by assuming a mean
strength of the magnetic field. This will be a key question in the
discussion of this paper.

The injection of nuclear particles from the suprathermal
tail of the momentum distribution produced in the dissipative shock
transition is much better understood,
because the same mechanism that produces
scattering fluctuations for higher-energy nuclei -- essentially a beam
instability from the accelerated nuclei that is so strong that
asymptotically the particles scatter along the mean field direction
already after one gyro-period -- also works at injection energies. In
addition, where heavy ion injection takes place
\citep{vbk03}, it is to be expected that these ions
dominate the nonthermal energy density behind a strong shock like in a
young SNR. Then the main nonlinear shock modification
consists in a weakening of the quasi-discontinuous part of the
shock structure, associated with a broad shock precursor. Low energy
particles -- ions and electrons -- in the accelerated spectrum are
then only accelerated at this weaker subshock and this implies a
significantly softer momentum spectrum at low energies than at high
energies. This physical effect is visible in the radio part of the
electron synchrotron spectrum and therefore a quantitative indication
of the degree of shock modification. It provides a means to determine
the injection rate of nuclear particles, de facto of protons, from the
radio synchrotron spectrum. In all cases, where the synchrotron
spectrum of SNRs was measured, this softening was observed. Together
with the nonlinear theory of acceleration, and in the strong
scattering limit, this determines the nonthermal pressure
$P_\mathrm{c}$, which turns out to be comparable with the kinetic
pressure $\rho V_\mathrm{s}^2$ of the gas. Here $ V_\mathrm{s}$ and
$\rho$ denote the shock velocity and the the upstream mass density,
respectively. Using therefore the synchrotron measurement, the
nonthermal quantities can be determined from theory.  The exception is
at first sight the mean magnetic field strength. However, it needs to
be consistent with the {\it overall form} of the synchrotron spectrum,
from radio to X-rays, and with the X-ray synchrotron {\it morphology}
that depends on the effective strength of the magnetic field. In this
way an interior effective magnetic field strength is determined. It is
typically an order of magnitude larger than the MHD-compressed
upstream field strength. This amplification of the magnetic
field is a characteristic of the effective acceleration of CR nuclei
in a SNR, because it can only be the result of strong acceleration of
nuclear particles. The pressure of the accelerated electrons --
also for the cases discussed below -- is more than two orders of
magnitude below $\rho V_\mathrm{s}^2$.

The question is then, whether the observed \gr emission is also
dominated by nuclear particles through their inelastic, $\pi^0$ -
producing collisions with thermal gas nuclei. This
need not be the case if the target density of the thermal gas is very
low, despite the fact that the energy density of the accelerated
nuclear particle component is very high, in fact comparable to the
thermal energy density.

Using this theoretical approach \citep[for reviews, see
e.g.][]{Malkov01,vbk04,ber05,ber08} the investigation of half a dozen of {\it
  young} Galactic SNRs has shown that the nuclear CR production is in all cases
so high that the Galactic SNRs are viable candidates for the Galactic CR
population up to particle energies $\sim 10^{17}$ eV, well above the so-called
knee in the spectrum \citep{bv07}. Even though important details are open to
debate because the time-dependent evolution of a point explosion can only be
calculated numerically \citep{byk96}, we believe that this result is quite a
robust one.

However, from a strictly observational point of view, the hadronic
nature of most of the SNR \gr sources is not proven this way. This
might ultimately be possible in a direct way with a very sensitive
neutrino detector. The remaining question whether the Galactic CR population has a SNR origin then
still requires the consistency of the observational result and the
theoretical picture.

As far as \gr observations are concerned, there is also a
different approach, basically phenomenological. It considers the
question, whether and to which extent the hadronic or leptonic origin
of the measured \gr emission can be decided by favoring either one
mechanism at the expense of the other directly from the data. For
example, it can ask the question whether the necessarily limited
dynamical range of the observed \gr emission allows a distinction
between a hadronic and a leptonic scenario. 
Or it can ask whether observations in other wavelength
ranges tend to empirically contradict the theoretically favored
scenario of a predominantly nuclear energetic particle energy
density. A possible topic consists in the interpretation of spatial
correlations in resolved \gr SNRs, like those noted in RX~J1713.7-3946
\citep{aha06} and RX~J0852.0-4622 (Vela Jr.) \citep{aha07}. The
correlation of the hard X-ray synchrotron emission with the VHE \gr
emission features might be considered to favor energetic electrons to
produce both emissions.  Discussions of the above and similar issues
have recently been given for instance in
\citet{aha06,port07,aha07,Katz08,Plaga08}, and \citet{bv08}. However,
the complexity of the configurations that characterize these extended
sources introduces severe uncertainties. They arise from the poorly
known structure of the circumstellar medium, which could be partly due
to the strong winds expected from the progenitor stars or could be
partly pre-existing in the form of neighboring interstellar
clouds, affected by the progenitor and its subsequent explosion.

We shall add in this paper such a phenomenological argument. It
concerns the spatially integrated synchrotron emission spectrum for
the simplest available objects, the remnants of the three young
type Ia SNe, observed in VHE $\gamma$-rays. Even though only upper limits exist
from the HEGRA, \hess \, and CANGAROO experiments for SN 1006
\citep{aha05}, Tycho's SNR \citep{aha01}, and Kepler's SNR
\citep{enomoto08,aha08}, they can nevertheless be used to estimate
lower limits to the effective mean magnetic field strengths in the SNR
that are consistent with the observed spatially-integrated synchrotron
spectra. These somewhat naively estimated magnetic fields are then
compared to the expectations for these types of
SN explosions. The large discrepancies found disfavor leptonic
scenarios for these objects.

\section{Simple synchrotron and IC modeling of the integrated 
emission}

The expected synchrotron spectral energy density (SED) at distance $d$
from a SNR is given by the expression \citep[e.g.][]{brz90}

\[
{E^2 {dF^\mathrm{syn}\over dE}}=
{3\times 10^{- 21}\over 4\pi d^2}
\int d^{3}r \, B_{\perp}
\]
\begin{equation}
\hspace{1.5cm} \times \int_0^{\infty} dp p^2 
f_\mathrm{e}(\vec{r},p) g\left(\frac{E} {h \nu_\mathrm{c}} \right)
\label{eq1}
\end{equation}
%

%
%
\noindent in erg/(cm$^2$s), where
\[
g(y)=y\int_y^{\infty}K_{5/3}(y')dy',
\]
$K_{\mu}(y)$ is the modified Bessel function, $E$ is the photon energy,
$\nu_\mathrm{c}=3e B_{\perp} p^2 /[ 4\pi (m_\mathrm{e}c)^3]$, 
and $B_{\perp}$ is the interior
magnetic field component perpendicular to the line of sight.

We shall use here an approximation that averages over the line of sight
directions. A precise analytical integration involving Whittacker's function
has been given by \citet{cru86} which is in turn closely approximated by
substituting
\begin{equation}
B_{\perp} = \sqrt{2/3} B_\mathrm{d},
\label{eq2}
\end{equation}
\noindent into Eq.(\ref{eq1}). Here $B_\mathrm{d}$ is the strength of the interior
field which results from the MHD-compression of the upstream magnetic field
$\vec{B_0}$ and subsequent de-compression in the interior (see below). The
strength of $\vec{B_0}$ is denoted as $B_0$.

The spatial integral in Eq.(\ref{eq1}) extends over the volume $V$ of the SNR, as
given by the observed synchrotron morphology, and the calculated synchrotron SED
has to be compared with the observed SED.

Our starting point for a simplified model is the assumption that $B_{\perp}$ in
the form of Eq.(\ref{eq2}) can be taken as a weighted mean value $\sqrt{2/3}
\langle B_\mathrm{d} \rangle $ out of the spatial integral of
Eq.(\ref{eq1}). The postshock value of $B_\mathrm{d}/B_0$ is locally between 1
and $\sigma$, where $\sigma > 1$ denotes the overall shock compression
ratio. Since the interior field strength is lower than the postshock field
strength, we have for this mean interior field strength: $\langle B_\mathrm{d}
\rangle < \sigma B_0 $.

If we investigate the possibility that the accelerated particles are electrons
alone -- implying a purely leptonic origin of the VHE \gr emission -- then we
have to consider a test particle problem with $\sigma =4$. In the same sense
$B_0$ should be equal to the strength of the interstellar magnetic field,
i.e. equal to a few $\mu$G. Values of $B_0=3 \mu$G and $B_0=5 \mu$G then imply
$3 < \langle B_\mathrm{d}\rangle < 12 \mu$G and $5 < \langle
B_\mathrm{d}\rangle < 20 \mu$G, respectively.

As a second approximation we shall also assume that the volume
integral of $f_\mathrm{e}(\vec{r},p)$ equals the product of $V$ and an
electron distribution
\begin{equation}
\int d^{3}r f_\mathrm{e}(\vec{r},p)= V A p^{-\alpha}\times
\exp(-p/p_\mathrm{max}),
\label{eq3}
\end{equation}
\noindent in the form of a power law with an exponential cutoff at
$p_\mathrm{max}$. The index $\alpha = 4$ again corresponds to a test particle
spectrum\footnote{For internal magnetic field strengths in excess of $100 \mu$G
  such a model distribution would have to include a high-energy part of the
  spectrum that is softened by synchrotron losses, see
  e.g. \citet{bkv02}. However in the present context, that does by assumption
  not include massive nuclear particle acceleration, such field strengths are
  not expected to occur.}.

In this sense the two parameters $A$ and $p_\mathrm{max}$ can be
approximately fitted from the known radio and X-ray synchrotron data
as a function of $B_0$. In fact, because of the exponential behaviour
of the cut-off of the electron momentum distribution the only
sensitive parameter turns out to be $A$. The fact that the
observed radio synchrotron spectra are softer than implied by a
distribution with $\alpha = 4$ suggests that the pure electron
acceleration model, for the sake of argument considered here, is not
the physically correct model. However, a pure electron model is
necessarily one with $\alpha = 4$, even if it does not optimally fit
the {\it form} of the observed synchrotron spectrum, but rather only
its {\it amplitude}.

Next we calculate the IC SED from these same electrons in the 2.7 K
cosmic microwave background (CMB). This \gr SED can be written in the
form:
\[
E^2 {dF_{\gamma}^\mathrm{IC}\over dE}=
{E^2 c \over d^2}
\int d^{3}r 
\int_0^{\infty} d\epsilon n_\mathrm{ph}(\epsilon)
\]
\begin{equation}
\hspace{1.5cm} \times \int_{p_\mathrm{min}}^{\infty} dp p^2
\sigma(\epsilon_\mathrm{e},E,\epsilon)f_\mathrm{e}(\vec{r},p) 
\label{eq4}
\end{equation}
\noindent in erg/(cm$^2$s), where \citep{blgo70}
\[
\sigma(\epsilon_\mathrm{e},E,\epsilon)=
\frac{3\sigma_\mathrm{T}(m_\mathrm{e}c^2)^2}{4\epsilon
\epsilon_\mathrm{e}^2}
\]
\begin{equation}
\hspace{1cm} \times \left[ 
2q\ln q+(1+2q)(1-q)+0.5\frac{(\Gamma q)^2(1-q)}{1+\Gamma  q}
\right]
\label{eq5}
\end{equation}
\noindent is the differential cross section for the up-scattering of a photon
with incident energy $\epsilon$ to energy $E$ by
the elastic collision with an electron of energy $\epsilon_\mathrm{e}$,
\begin{equation}
n_\mathrm{ph}=\frac{1}{\pi^2 (\-\hbar c)^3}
\frac{\epsilon ^2}{\exp(\epsilon/k_\mathrm{B}T)-1}
\label{eq6}
\end{equation}
\noindent is the blackbody spectrum of the CMB, $h=2\pi\hbar$ and $k_B$ are the
Planck and Boltzmann constants, respectively, $T=2.7$~K,
$\sigma_\mathrm{T}=6.65 \times 10^{-25}$ cm$^2$ is the Thomson cross-section,
$q=E/[\Gamma (\epsilon_\mathrm{e} -E)]$, $\Gamma = 4\epsilon
\epsilon_\mathrm{e} /(m_\mathrm{e}c^2)^2$ , and $p_\mathrm{min}$ is the minimal
momentum of the electrons, whose energy $\epsilon_\mathrm{e}$ is determined by the
condition $q=1$.

We neglect here nonthermal Bremsstrahlung emission which turns out to
be unimportant for all the cases considered below.

\begin{figure*} 
\centering 
\includegraphics[width=0.8\textwidth]{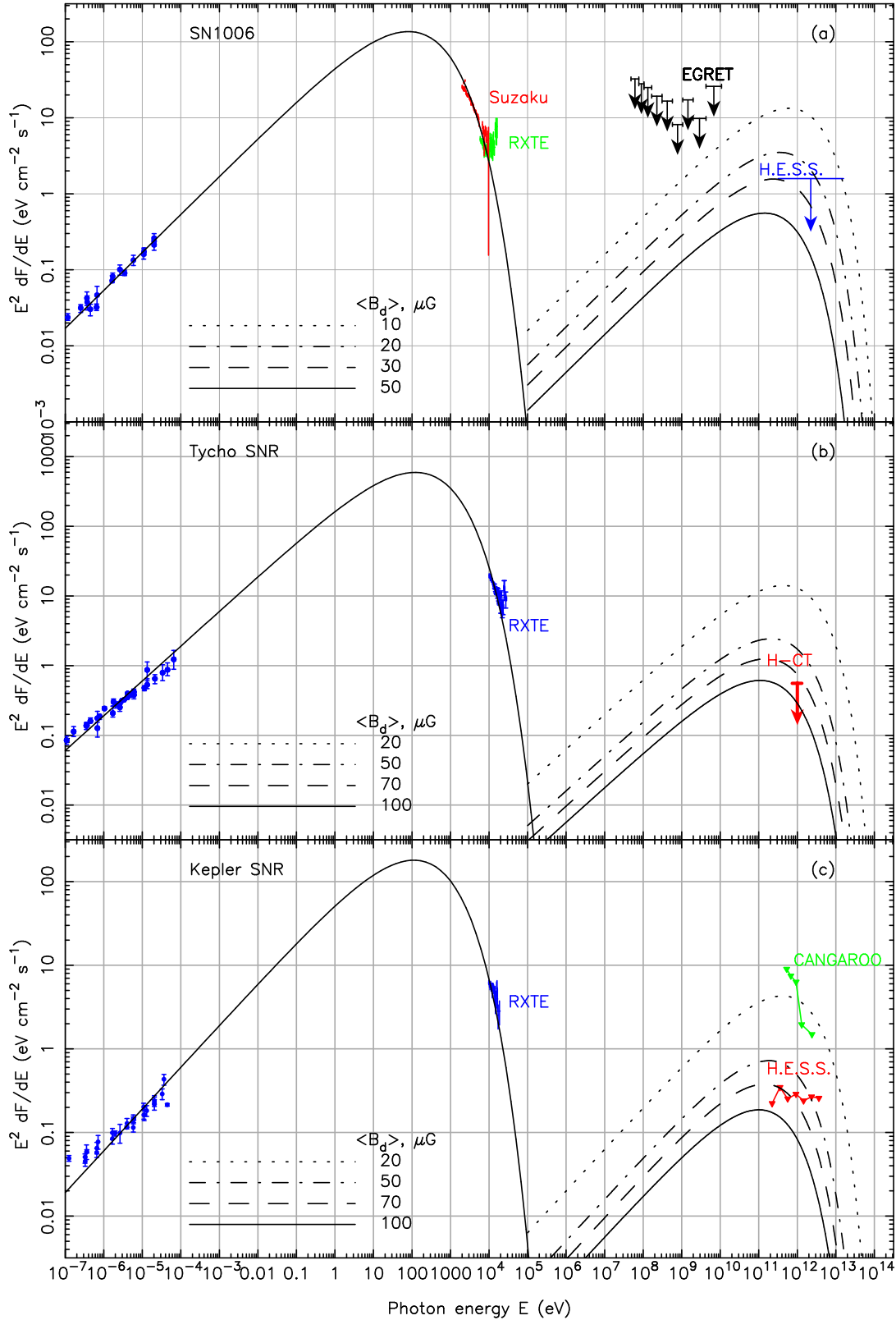}
\caption{The overall (spatially integrated) nonthermal spectral energy
  distribution (SED) as a function of photon energy $E$. The lower-energy part
  shows the simple fit to the observed synchrotron-SED, cf. Eq.(\ref{eq3}), for
  various values of the internal field strength $B_{\perp}$ in $\mu$G,
  cf. Eq.(\ref{eq2}). The synchrotron fit is essentially independent of mean
  strength $\langle B_\mathrm{d} \rangle$ of the internal field. The
  high-energy curves show the inverse Compton-SED in the CMB for the various
  field strengths. {\em (a)} for SN1006: The {\it blue} radio data are from
  \citet{rey96} whereas the {\it green} and {\it red} X-ray data are from RXTE
  \citep{allen99}, and Suzaku \citep{bfh08}, respectively. The
  Chandra data \citep{allen04} are very similar to the Suzaku data and can be
  treated as indistinguishable in the present context. Also given are the upper
  limits from \hess \citep{aha05} and EGRET \citep{naito99}.
  (b) for Tycho's SNR: The radio data (in {\it
    blue}) are from \citet{re92}, whereas the X-ray data (in {\it blue}) are
  from RXTE \citep{allen99}. The \gr upper limit (in {\it red}) is from the
  HEGRA Cherenkov telescope (H-CT) system \citep{aha01}.  (c) for Kepler's
  SNR: The radio data (in {\it blue}) are from \citet{re92}, whereas the X-ray
  data (in {\it blue}) are again from RXTE \citep{allen99}. The \gr upper
  limits are from CANGAROO \citep{enomoto08} (in {\it green}) and from \hess
  \citep{aha08} (in {\it red}).}
\label{f1} 
\end{figure*}

Since the CMB is uniform, we can without further approximation use
Eq.(\ref{eq3}) to express the \gr SED in terms of the parameters $A$ and
$p_\mathrm{max}$. The results are given in Fig.~\ref{f1}a for SN 1006, in 
Fig.~\ref{f1}b for Tycho's and in Fig.~\ref{f1}c for Kepler's SNRs for 
various values of $\langle B_\mathrm{d}\rangle$. 

\section{Discussion}

It is clear from the outset that the approximate nature of the models
used limits the impact of the conclusions to be drawn from these
results. On the other hand, most of the arguments that have been used
in the past regarding the alternative between a hadronic and a
leptonic interpretation of VHE \gr results have used such one box
approximations. The only alternative would be full time-dependent
solutions of the governing system of equations, discussed in the
Introduction.

However, with this proviso, the results are surprisingly clear. For
all three sources magnetic field strengths $B_0$ lower or equal to the
expected interstellar magnetic fields of $3-5 \mu$G substantially
overpredict even the existing \gr upper limits.

For a shock that excites MHD fluctuations only weakly if at all, because of the
assumed lack of acceleration of nuclear particles, the interior gas flow will
be essentially laminar and adiabatic. Taking into account that in such an
approximately laminar gas flow the minimum strength of the internal magnetic
field will always be lower than the strength of the upstream field and that the
weighted average field strength average field strength is considerably lower
than the maximum field strength (over the quasi-circular shock surface)
immediately behind the shock, a more realistic estimate for the value of
$\langle B_\mathrm{d}\rangle$ would be to put $\sigma \sim 1$. This would imply
that the $\langle B_\mathrm{d}\rangle$\,-\,values, given in Fig.\ref{f1},
roughly equal the values of $B_0$. This means that already the curves for
$B_\mathrm{d}= 10\mu$G in the figures assume an unrealistically high ambient
interstellar field strength, larger than the interstellar average. Yet they
overpredict the IC \gr flux by at least one order of magnitude in comparison
with the observed total \gr {\it upper limit} already for the very low-density
object SN 1006 \citep{Acero07} -- that would therefore be expected to be
located in a lower than average interstellar magnetic field as well -- and by
much more for the two other sources.

Existing theoretical solutions for the overall particle acceleration in these
three sources take into account the amplification of the magnetic field by the
accelerating nuclear particles whose energy density becomes comparable to the
kinetic energy of the incoming gas flow, as seen in the frame of the shock
\citep[e.g.][]{kbv05,vbk05a,bkv06}.  Only then it seems possible to not
overpredict the leptonic flux. At the same time the \gr flux is dominated by
the hadronic flux, even though in SN 1006 only by a small margin.

We note here that besides nonlinear amplification due to CRs the magnetic field
in SNRs can also by amplified by other mechanisms. These are Rayleigh-Taylor
instabilities at the contact discontinuity between the ejecta and the shocked
circumstellar gas \citep[e.g.][]{wang01} and the vorticity generation that
results from the shock running into possibly pre-existing density
inhomogeneities of the circumstellar medium \citep{gj07}.  However, the common
feature of these mechanisms is that they act only in the downstream region and
produce their main effect at a substantial distance behind the shock, that is
outside the CR acceleration region. Therefore these mechanisms do not influence
the CR acceleration process, in particular the maximal particle energy.  Since
accelerated CRs in young SNRs are concentrated in a thin layer near the shock
front, these two mechanisms also hardly influence the properties of the
nonthermal emission, produced by CRs, except possibly that of the
highest-energy CRs.

\section{Conclusions}

Simple one box approximations indicate that a leptonic scenario for
the \gr emission from the three known Galactic type Ia SNRs SN 1006,
Tycho's SNR and Kepler's SNR significantly overpredicts the \gr flux,
even when compared to the existing upper limits from observations. The
calculation makes direct use of the observed synchrotron emission
spectra. Even though the arguments are simplistic, they appear to
eliminate equally simplistic phenomenological arguments in favor of
such a scenario. Any positive argument in favor of a purely
leptonic scenario would therefore have to be based on a full solution
of the governing nonlinear equations. From our results, however, we
believe that such a positive argument can not be made.

\begin{acknowledgements}
We are indebted to Drs. Glen Allen and Aya Bamba for providing us the X-ray
spectra for SN 1006 from Chandra and Suzaku in physical units. This work has
been supported in part by the Russian Foundation for Basic Research (grants
06-02-96008, 07-02-0221).  EGB and LTK acknowledge the hospitality of the
Max-Planck-Institut f\"ur Kernphysik, where part of this work was carried
out.
\end{acknowledgements}

\end{document}